\documentclass[prl,twocolumn]{revtex4}%
\usepackage{amssymb}
\usepackage{graphicx}
\usepackage{amsmath}
\usepackage{amsfonts}%
\begin{document}
\title{An electron spin injection driven paramagnetic solid state MASER device}
\author{S. M. Watts}
\author{B. J. van Wees}
\affiliation{Physics of Nanodevices, Materials Science Centre, University of Groningen,
Nijenborgh 4, 9747 AG Groningen, The Netherlands}
\date{\today }

\begin{abstract}
In response to an external, microwave-frequency magnetic field, a paramagnetic
medium will absorb energy from the field that drives the magnetization
dynamics. \ Here we describe a new process by which an external spin injection
source, when combined with the microwave field spin-pumping, can drive the
paramagnetic medium from one that absorbs microwave energy to one that emits
microwave energy. We derive a simple condition for the crossover from
absorptive to emissive behavior. \ Based on this process, we propose a spin
injection-driven paramagnetic MASER device.
\end{abstract}
\maketitle

In spin electronics (\textquotedblleft spintronics\textquotedblright), the
spin degree of freedom of the electron is used as a new means of storing,
manipulating and transferring signals and information \cite{wolf}.  Since the
prediction that a spin-polarized current can excite the (stimulated) emission
of spin waves at a ferromagnetic interface \cite{berger}, it has been
demonstrated in magnetic multilayer systems that spin-polarized currents can
drive microwave-frequency magnetization dynamics in the ferromagnetic layers
\cite{tsoi,kiselev,rippard}. \ More recently, it has been proposed that the
magnetization dynamics of a ferromagnet can be used as a \textquotedblleft
spin pump\textquotedblright\ to drive a spin current into an adjacent,
non-magnetic metallic layer \cite{brataas}. \ In all of the above, the
dynamics of the magnetization in the ferromagnetic layer is of primary importance.

We have recently shown, however, that a ferromagnet is not required for spin
pumping, since applied radio-frequency (rf) magnetic fields can produce spin
accumulation in paramagnetic materials \cite{watts}. In this case, the
paramagnetic medium absorbs energy from the external microwave magnetic field
to produce the non-equilibrium spin accumulation. \ In this letter we describe
a new effect in which the energy flow can be reversed: by combining an
external spin injection source with microwave field spin pumping in a
paramagnetic medium, the system can be driven from one that absorbs microwave
energy to one that emits microwave energy. \ When the medium is placed within
a resonant circuit, it is possible to obtain MASER (Microwave Amplification by
Stimulated Emission of Radiation) action driven by spin injection.

The operation of a MASER device requires the generation of population
inversion. \ In a two-level paramagnetic MASER system, both excitation and
stimulated emission occur from the same pair of levels \cite{fain}.
\ Population inversion can be produced by separating the processes of
excitation and emission in time, for instance by pulsed excitation. In
multi-level systems the levels used for excitation and emission can be
separated, and continuous operation is then possible \cite{fain}. \ For
example, a proton nuclear spin MASER was realized using the coupling between
the nuclear and electron spins, and exciting the latter with microwaves
\cite{abragam}, or by optically pumping nuclear transitions in $^{3}$He
\cite{robinson}. \ The continuing progress in the field of spintronics in both
technological advances as well as the understanding of fundamental processes,
offers a new possibility in which a spin current injected into the
paramagnetic medium can act as a continuous source of population
inversion.\ \ As we will show, this allows for a straightforward description
of MASER\ operation with semi-classical Bloch-type equations written in terms
of spin accumulation. \ The proposed spin injection driven MASER, at least in
its present form, is not a practical design for use in applications, but we
show that it is feasible to fabricate and study one with the current level of
technology, as demonstrated in, for instance, spin current switching
experiments \cite{spincurrent} and microwave pulsed field-induced
ferromagnetic switching \cite{switching}.
\begin{figure}[ptb]
\begin{center}
\includegraphics[width=8cm]{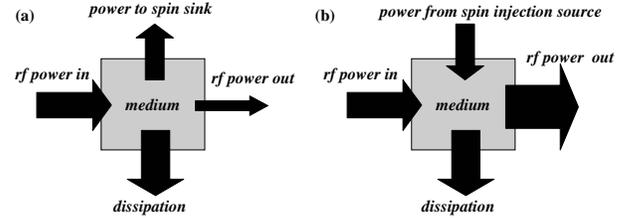}%
\caption{Power flow in and out of a paramagnetic medium to illustrate the
principle behind the spin injection MASER. (a) Part of the input rf power is
absorbed to produce precessing spin accumulation in the medium; power is lost
to internal dissipation due to spin relaxation processes, and to a spin sink.
(b) When additional spins are provided by an externally-powered spin injection
source, it is possible for the system to exhibit gain, resulting in net
emission of rf power.}%
\label{power}%
\end{center}
\end{figure}

\begin{figure}[ptb]
\begin{center}
\includegraphics[width=7cm]{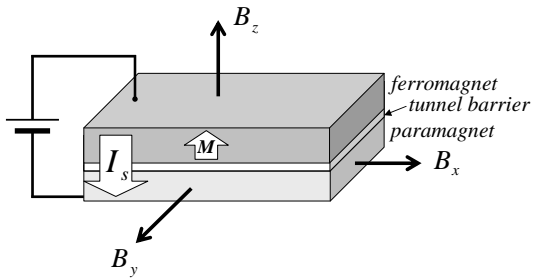}
\caption{Schematic of an electrical spin injection source for the paramagnetic
medium. An electrical current is driven across the interface between a
ferromagnetic layer and the paramagnetic layer (separated by an oxide tunnel
barrier), which causes a spin current $I_{s}$ to flow into the paramagnet. It
is assumed that the static field $B_{z}$ is sufficiently strong to saturate
the magnetization $M$ of the ferromagnetic layer normal to the film-plane.}%
\label{system}
\end{center}
\end{figure}

The principle behind the device is illustrated in Fig. \ref{power}. In Fig.
\ref{power}a, part of the rf power incident on a paramagnetic medium is
absorbed in the process of creating spin accumulation in the system. Power is
lost to internal dissipation due to spin relaxation and to an external
\textquotedblleft spin sink.\textquotedblright\ \ In a real device this could
be a ferromagnetic electrode used to probe the spin accumulation in the
paramagnetic region. \ However, the role of the spin sink can be reversed by
driving a spin-polarized current into the paramagnet from an
externally-powered spin source, as shown in Fig. \ref{power}b. \ As we will
demonstrate below, this allows the medium to exhibit rf power gain. \ Fig.
\ref{system} shows schematically a bilayer structure for injecting spins
electrically: a ferromagnetic layer is deposited on top of the paramagnetic
layer (separated by a thin aluminum oxide tunnel barrier), and the layers are
connected to opposite terminals of a voltage source so that an electrical
current is driven across the interface between the layers. This will cause a
spin current $I_{s}$ to flow into the paramagnetic layer, whose thickness is
assumed to be much less than its spin relaxation length so that a uniform spin
accumulation $\mu_{s}$ will build up in the layer. \ The role of the
ferromagnet in Fig. \ref{system} is purely as a source of spin current. \ The
dynamics of the magnetization or spin waves in the ferromagnet are not
relevant to our description. \ We will first show how the injected spin
current modifies the solutions for the spin-pumped spin accumulation in the
medium, and then analyze the response of the system within an electrical
circuit model.

In a previous publication \cite{watts}, we have described the dynamics of the
spin accumulation $\vec{\mu}=(\mu_{x},\mu_{y},\mu_{z})$ in response to a
time-dependent magnetic field $\vec{B}(t)$. We now extend this description to
include interaction with an external spin source:%
\begin{equation}
\frac{d}{dt}\vec{\mu}(t)=-\hbar\frac{d}{dt}\vec{\omega}_{B}(t)-\frac{\vec{\mu
}(t)}{\tau}+\vec{\omega}_{B}(t)\times\vec{\mu}(t)+\vec{I}_{s}(t),
\label{Pump1}%
\end{equation}
where $\vec{I}_{s}(t)$ is the injected spin current (in units of power, as
described below), $\vec{\omega}_{B}$ is the Larmor frequency $\hbar\vec
{\omega}_{B}=g\mu_{B}\vec{B}$ ($g\simeq2$ is the electron g-factor), and
$\tau$ is the spin relaxation time. \ Eq. \ref{Pump1} describes the time
evolution of the spin accumulation in response to four physical processes (the
four terms on the right-hand side of the equation): \ \textquotedblleft spin
pumping,\textquotedblright\ spin relaxation, spin precession, and spin
injection. \ The spin pumping term describes the rate at which the
time-dependent magnetic field, via the Zeeman energy, pushes spins aligned
with the magnetic field below the Fermi level, and anti-aligned spins above
the Fermi level, behaving essentially as a source of locally-injected,
time-dependent spin currents \cite{watts}. \ In a uniform, paramagnetic
system, we previously found that for a magnetic field rotating with angular
frequency $\omega$ in the x-y plane, a constant spin accumulation is generated
in the z-direction. \ Eqn. \ref{Pump1} (with $\vec{I}_{s}(t)=0$) may be
connected to electron spin resonance theory by observing that the
time-dependent magnetization is given by $\vec{m}(t)=\frac{1}{4}gN_{F}\mu
_{B}\left(  \vec{\mu}(t)+\hbar\vec{\omega}_{B}(t)\right)  $, where $N_{F}$ is
the total electron density of states at the Fermi level. The first term
$\vec{\mu}(t)$ in the expression for $\vec{m}(t)$ describes the
non-equilibrium magnetization, and the second term describes the equilibrium
magnetization due to the field-induced paramagnetism \cite{whitfield,abragam}.

\begin{figure}[ptb]
\begin{center}
\includegraphics[width=8cm]{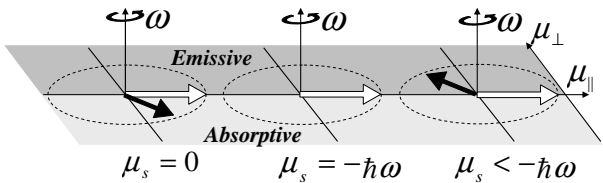}
\caption{A vector diagram in the rotating reference frame showing the field
(white arrow) and the in-plane spin accumulation (black arrow), for three values
of $\mu_{s}$. The $\mu_{\Vert}$ axis is the spin accumulation in-phase with
the field, while the $\mu_{\bot}$ axis is the spin accumulation out-of-phase
with the field. The half plane $\mu_{\bot}<0$ represents in-plane $\mu$
that lags the field (absorption), the half-plane $\mu_{\bot}>0$
represents in-plane $\mu$ that leads the field (emission).}%
\label{RotFrame}
\end{center}
\end{figure}

The system in Fig. \ref{system} is now spin-pumped by applying a magnetic
field $B_{xy}$ that rotates in the x-y plane at microwave frequency $\omega$,
and a static field $B_{z}$ is applied in the z-direction; $\omega_{xy}$ and
$\omega_{z}$ are the Larmor frequencies associated with these fields. A
constant spin current $\vec{I}_{s}(t)=I_{s}\hat{z}=\mu_{s}\tau^{-1}\hat{z}$ is
injected into the system, where $\mu_{s}$ is the spin accumulation that would
be present in the paramagnet in the absence of the applied field, and is
collinear with $B_{z}$. \ The sign of $\mu_{s}$ depends on the orientation of
the magnetization (usually fixed by $B_{z}$) and the direction of charge
current flow across the interface. The spin accumulation is found by solving
Eq. \ref{Pump1} after transforming to a rotating reference frame in which the
in-plane magnetic field is static \cite{watts}. The solutions are:
\begin{align}
\mu_{\Vert}  &  =\frac{(\omega_{z}-\omega)\tau(\omega_{xy}\tau)}%
{1+(\omega_{xy}\tau)^{2}+(\omega_{z}-\omega)^{2}\tau^{2}}(\hbar\omega+\mu
_{s})\label{ull}\\
\mu_{\bot}  &  =-\frac{(\omega_{xy}\tau)}{1+(\omega_{xy}\tau)^{2}+(\omega
_{z}-\omega)^{2}\tau^{2}}(\hbar\omega+\mu_{s})\label{up}\\
\mu_{z}  &  =\mu_{s}-\frac{(\omega_{xy}\tau)^{2}}{1+(\omega_{xy}\tau
)^{2}+(\omega_{z}-\omega)^{2}\tau^{2}}(\hbar\omega+\mu_{s}), \label{uz}%
\end{align}
where $\mu_{\Vert}$ and $\mu_{\bot}$ refer to the components of the spin
accumulation that rotate in-phase (dispersive component) and $90^{\circ}$
out-of-phase (absorptive/emissive component) with the field $B_{xy}$,
respectively. \ In our previous work we focused on $\mu_{z}$, which can be
measured electrically in a dc transport experiment by using ferromagnetic
detection electrodes. \ Here, with the addition of the injected spin current,
we instead focus on $\mu_{\Vert}$ and $\mu_{\bot}$. In Fig. \ref{RotFrame} we
illustrate three cases for specific values of $\mu_{s}$. \ When $\mu_{s}=0$,
$\mu_{\bot}$ is always negative. This situation is shown in Fig.
\ref{RotFrame}a, where the black vector is the in-plane spin accumulation in
the rotating reference frame. The vector will always lie in the half-plane
$\mu_{\bot}<0$, which represents energy absorption from the field. When
$\mu_{s}=-\hbar\omega$, we have that $\mu_{\Vert}=\mu_{\bot}=0$ and $\mu
_{z}=\mu_{s}$, thus giving the remarkable result that the spin pumping effect
can be turned off with an injected spin current. This is shown in
Fig.\ref{RotFrame}b, where the in-plane spin accumulation vector has
disappeared. When $\mu_{s}<-\hbar\omega$, both $\mu_{\Vert}$ and $\mu_{\bot}$
change sign as shown in Fig. \ref{RotFrame}c. The spin accumulation vector now
lies in the half-plane $\mu_{\bot}>0$, which represents energy emission to the
field. \ In MASER (or laser) terminology, this corresponds to stimulated
emission resulting in the medium exhibiting gain.

In other words: \ the injected spins, originally oriented along the
z-direction, will precess around the field $B_{xy}$. \ Depending on their
initial polarization in the z-direction, they will give either a positive or
negative contribution to $\mu_{\bot}$, and depending on the sign of $\omega$,
they will either enhance the damping or generate gain.

Using Eqs. \ref{ull}-\ref{uz} we find the power per unit volume in the
paramagnet dissipated due to spin relaxation:%
\begin{equation}
P_{diss}=\frac{N_{F}}{4\tau}|\vec{\mu}|^{2}=\frac{1}{4\tau}N_{F}(\mu_{s}%
^{2}-\omega_{xy}\tau(\hbar\omega-\mu_{s})\mu_{\bot}). \label{Pdiss}%
\end{equation}
The net power extracted from or supplied to the field is determined by the
time-averaged value of $\vec{m}\cdot\frac{d}{dt}\vec{B}$ \cite{landau}:%
\begin{equation}
P_{field}=-\frac{1}{4\tau}N_{F}\omega_{xy}\tau\mu_{\bot}\hbar\omega.
\label{Pfield}%
\end{equation}
This power can be positive or negative, as discussed above. \ The power
supplied from the external spin source is
\begin{equation}
P_{in}=\frac{1}{4}I_{s}\mu_{z}N_{F}=\frac{1}{4\tau}N_{F}(\mu_{s}^{2}%
+\omega_{xy}\tau\mu_{s}\mu_{\bot}).
\end{equation}
$P_{in}+P_{field}=P_{diss}$, as required by conservation of energy
\cite{powernote}.

While it is possible to generate locally rotating rf magnetic fields in a
device, it is generally more practical to use a linear rf driving field in
resonance with $\omega_{z}$, similar to conventional elecron spin resonance
experiments. A linear field can be decomposed into two counter-rotating
magnetic fields of half the magnitude. \ For positive $\omega_{z}$, only the
field rotating counter-clockwise has the right sense for the resonant
condition $\omega=\omega_{z}$ to hold, and thus to contribute significantly to
the response. \ We have checked this explicitly for the relevant range of
parameters with exact, numerical solutions of Eq. \ref{Pump1}. \ To simplify
the expressions that follow, we will therefore assume the resonant condition
with the convention of positive $\omega$, and further assume that
$(\omega_{xy}\tau)^{2}<<1$, which is typically the case experimentally. We
then have $\mu_{\Vert}\simeq0$ and $\mu_{\bot}\simeq-\omega_{xy}\tau
(\hbar\omega+\mu_{s})$. \ The in-plane magnetization components $m_{\Vert
}=\frac{1}{4}gN_{F}\mu_{B}\hbar\omega_{xy}$ \ and $m_{\bot}=-\frac{1}{4}%
gN_{F}\mu_{B}\omega_{xy}\tau(\hbar\omega+\mu_{s})$ are now both linear in
$\omega_{xy}$, and we can define a linear, complex magnetic susceptibility
$\chi=\chi_{\Vert}+i\chi_{\bot}$ for $\mu_{0}(m_{\Vert}+im_{\bot})=\chi
B_{xy}$. \ Then the components of the susceptibility are:%
\begin{align}
\chi_{\Vert}  &  =\frac{1}{4}gN_{F}\mu_{B}^{2}\mu_{0}\equiv\chi_{0}\\
\chi_{\bot}  &  =-\chi_{0}\frac{\tau}{\hbar}(\hbar\omega+\mu_{s}),
\end{align}
where $\chi_{0}$ is a constant determined by the material parameters $g$ and
$N_{F}$.
\begin{figure}[ptb]
\begin{center}
\includegraphics[width=8cm]{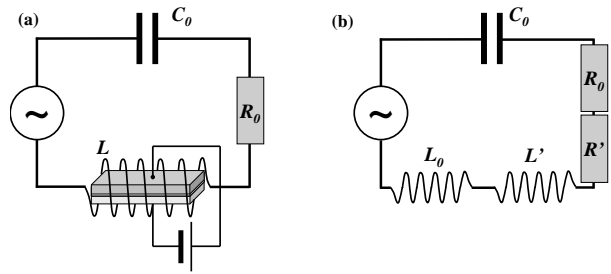}
\caption{(a) Equivalent LRC circuit for the spin-injection MASER connected to
an rf source. The paramagnetic medium is enclosed within the inductor coil
(the bare coil has inductance $L_{0}$). \ We assume that the capacitance
$C_{0}$ is chosen or tuned such that the resonant frequency of the circuit
$\omega_{0}^{2}=\left(  R_{0}C_{0}\right)  ^{-1}$ is matched with $\omega$.
(b) The magnetic susceptibility of the paramagnet modifies the inductance to
produce an additional effective circuit resistance $R^{\prime}$ and inductance
$L^{\prime}$. $R^{\prime}$ can be negative due to the spin injection MASER
effect.}
\label{circuit}
\end{center}
\end{figure}

In order to analyze the performance of our system within a device, we model
the system as the LRC circuit shown in Fig. \ref{circuit}. We follow a
standard treatment for spin resonance theory \cite{abragam,yukalov}. The
inductor coil is used to both excite and detect the magnetization induced in
the medium. The inductance is modified by $\chi$ as $L=L_{0}(1+\eta\chi)$,
where $\eta$ is the filling factor of the medium in the coil
\cite{fillingfactor}. The total impedance of the circuit is now given by
\begin{align}
Z &  =R_{0}+i\omega L+\frac{1}{i\omega C_{0}}\label{impedance}\\
&  =R_{0}-\omega L_{0}\eta\chi_{\bot}+i\omega L_{0}(1+\eta\chi_{\Vert}%
)+\frac{1}{i\omega C_{0}}.\nonumber
\end{align}
$\chi_{\bot}$ contributes to an additional resistance in the circuit,
$R^{\prime}=-\omega L_{0}\eta\chi_{\bot}=\omega L_{0}\eta\chi_{0}\frac{\tau
}{\hbar}\left(  \hbar\omega+\mu_{s}\right)  $ which can be positive or
negative. \ With $\mu_{s}=0$, $\mu_{\bot}$ will always be negative
(absorptive) and therefore so will $\chi_{\bot}$. $R^{\prime}$ is then
positive and always increases the damping in the circuit. But with $\mu
_{s}<-\hbar\omega$, both $\mu_{\bot}$ and $\chi_{\bot}$ are positive, and
$R^{\prime}$ is therefore negative. The reduction in the damping can be
measured as a change in the transmitted power in a suitably designed waveguide
structure \cite{grundler}. \ Self-sustaining, MASER oscillations
\cite{vanderPol} are obtained when the magnitude of the negative $R^{\prime}$
is larger than $R_{0}$, leading to negative effective resistance in the
circuit. We can write the condition $R_{0}+R^{\prime}<0$ in terms of the
quality factor $Q=\frac{\omega L_{0}}{R_{0}}$ of the unloaded circuit:
\begin{equation}
Q>\left(  -\eta\chi_{0}\frac{\tau}{\hbar}(\hbar\omega+\mu_{s})\right)  ^{-1}.
\end{equation}

Using typical parameters for Al with $\tau$ of order $10^{-10}$ s at room
temperature, we find that the product $\eta\chi_{0}\frac{\tau}{\hbar}$ is of
order 1/eV. \ Thus, for $\mu_{s}<<-\hbar\omega$ we have, effectively,
$Q>1/|\mu_{s}(eV)|$. \ To find a value for $Q$ we then only need to estimate
$\mu_{s}$. \ Spin relaxation in a confined medium can be described as an
effective resistance between parallel spin up and spin down channels. \ For
injection into an Al metal island ($500\times500\times50$ nm$^{3}$), a spin
resistance of 10 $\Omega$ was found experimentally \cite{zaffalon}. Scaling to
a larger medium $1\times1\times0.2\mu$m$^{3}$, the spin resistance is reduced
to about 1 $\Omega$. \ Assuming a ferromagnet polarization of 50\% and a
tunnel barrier resistance of 1 k$\Omega$ (about the lowest value that can be
obtained technologically) that limits the charge current to 10 mA, with which
a spin accumulation voltage $\mu_{s}\ $of order 5 meV should be achievable
(this is indeed much greater than $\hbar\omega$ for microwave frequencies up
to tens of GHz). \ Thus we estimate that $Q$ should be greater than 200.
\ Although this is a relatively large $Q$, it should be possible to fabricate
such an on-chip resonator, especially in a proof-of-principle experiment in
which superconducting strip lines are used to construct the resonant circuit
\cite{superconductor}. \ Other sources of dissipation should also be
considered, such as eddy-current damping. \ The planar, thin-film structure
shown in Fig. \ref{system} has the advantage that the fields $B_{x}$ and
$B_{y}$ oscillate in the film plane, thus reducing the importance of this effect.

To estimate the typical microwave power which can be emitted by this device,
we assume that the amplitude of the oscillating $B_{xy}$ is limited by the
condition $\omega_{xy}\tau\lesssim1$, at which point the stimulated emission
becomes less efficient (Eqs. \ref{up} and \ref{Pfield}). \ From Eq.
\ref{Pfield} we then find that a typical power is 1 $\mu$W (note, however,
that this scales with the device dimensions). \ The total power supplied by
the voltage source is of order 100 mW. \ The main reason for this low
efficiency is the mismatch between the resistance of the tunnel barrier (1
k$\Omega$) and the typical \textquotedblleft spin resistance\textquotedblright%
\ (1 $\Omega$) of the system. \ This is purely a technological problem related
to the limitations of using aluminum oxide tunnel barriers. \ 

In conclusion, we have shown that by combining an external spin injection
source with microwave field spin pumping on a paramagnetic medium, it can be
made to exhibit microwave energy gain for sufficiently large injected spin
currents. \ This effect does not depend on the details of the paramagnet or
the spin injection source. \ While we have used a simple metallic multilayer
system to illustrate our proposal for a spin-injection MASER, our description
can be readily applied to other systems, such as semiconductors, and other
sources of spin injection, such as optical pumping.

We acknowledge useful discussions with C. H. van der Wal and W. van Roy.
\ Support for this work has been provided by the Stichting Fundamenteel
Onderzoek der Materie (FOM).


\begin{thebibliography}{99}                                                                                         

\bibitem {wolf}S. A. Wolf, D. D. Awschalom, R. A. Buhrman, J. M. Daughton, S.
von Moln\'{a}r, M. L. Roukes, A. Y. Chtchelkanova, and D. M. Treger, Science
\textbf{294}, 1488 (2001).

\bibitem {berger}L. Berger, Phys. Rev. B \textbf{54}, 9353 (1996).

\bibitem {tsoi}M. Tsoi, A. G. M. Jansen, J. Bass, W.-C. Chiang, V. Tsoi and P.
Wyder, Nature \textbf{406}, 46 (2000).

\bibitem {kiselev}S. I. Kiselev, J. C. Sankey, I. N. Krivorotov, N. C. Emley,
R. J. Schoelkopf, R. A. Buhrman and D. C. Ralph, Nature \textbf{425}, 380 (2003).

\bibitem {rippard}W. H. Rippard, M. R. Pufall, S. Kaka, S. E. Russek, and T.
J. Silva, Phys. Rev. Lett. \textbf{92}, 027201 (2004).

\bibitem {brataas}A. Brataas, Y. Tserkovnyak, G. E. W. Bauer, and B. I.
Halperin, Phys. Rev. B \textbf{66}, 060404(R) (2002).

\bibitem {watts}S. M. Watts, J. Grollier, C. H. van der Wal, and B. J. van
Wees, Phys. Rev. Lett. \textbf{96}, 077201 (2006).

\bibitem {fain}V. M. Fain and Ya. I. Khanin, Quantum Electronics Vol. 2 (1969).

\bibitem {abragam}A. Abragam, The Principles of Nuclear Magnetism (1961).

\bibitem {robinson}H. G. Robinson and T. Myint, Appl. Phys. Lett. \textbf{5},
116 (1964).

\bibitem {spincurrent}E.B. Myers, D.C. Ralph, J.A. Katine, R.N. Louie, and
R.A. Buhrman, Science \textbf{285}, 867 (1999); J.A. Katine, F.J. Albert, R.A.
Buhrman, E.B. Myers, and D.C. Ralph, Phys. Rev. Lett. 84, 3149 (2000).

\bibitem {switching}Th. Gerrits, H. A. M. van den Berg, J. Hohlfeld, L.
B\"{a}r, and Th. Rasing, Nature \textbf{418}, 509 (2002); H.W. Schumacher, C.
Chappert, P. Crozat, R. C. Sousa, P. P. Freitas, J. Miltat, J. Fassbender, and
B. Hillebrands, Phys. Rev. Lett. \textbf{90}, 017201 (2003).

\bibitem {whitfield}G. Whitfield and A. G. Redfield, Phys. Rev. \textbf{106},
918 (1957).

\bibitem {landau}L. D. Landau and E. M. Lifshitz, Electrodynamics of
Continuous Media (1960).

\bibitem {powernote}In addition to these power flows related to the spin
degree of freedom, the external voltage source also injects a substantial
amount of power unrelated to spin. \ This will be dissipated as heat, and will
not affect the system.

\bibitem {yukalov}A more detailed treatment of the coupling between a spin
sample and the resonator can be found in V. I. Yukalov, Laser Physics
\textbf{5}, 970 (1995).

\bibitem {fillingfactor}We will assume $\eta$ is of order 1. \ An integral
expression for the filling factor is given by N. Bloembergen and R. V. Pound,
Phys. Rev. \textbf{95}, 8 (1954).

\bibitem {grundler}E.g., see F. Giesen, J. Podbielski, T. Korn, and D.
Grundler, J. Appl. Phys. \textbf{97}, 10A712 (2005).

\bibitem {vanderPol}Negative resistance oscillators are usually referred to as
van der Pol oscillators, see B. van der Pol, Phil. Mag. Series 7, \textbf{3},
65 (1927).

\bibitem {zaffalon}M. Zaffalon and B. J. van Wees, Phys. Rev. Lett.
\textbf{91}, 186601 (2003).

\bibitem {superconductor}A. Wallraff, D. Schuster, A. Blais, L. Frunzio, R.-S.
Huang, J. Majer, S. Kumar, S. M. Girvin, and R. J. Schoelkopf, cond-mat/0407325.
\end{thebibliography}
\end{document}